\documentclass[prd,twocolumn,showpacs]{revtex4}
\usepackage{epsfig}
\usepackage{graphicx}
\usepackage{amsmath}
\usepackage{color}

\newcommand{\comment}[1]{}

\newcommand\lsim{\mathrel{\rlap{\lower4pt\hbox{\hskip1pt$\sim$}}
        \raise1pt\hbox{$<$}}}
\newcommand\gsim{\mathrel{\rlap{\lower4pt\hbox{\hskip1pt$\sim$}}
        \raise1pt\hbox{$>$}}}

\newcommand{\mH}{m_{\rm H}}
\newcommand{\nH}{n_{\rm H}}
\newcommand{\Tr}{T_{\rm r}}
\newcommand{\Tm}{T_{\rm m}}
\newcommand{\nula}{\nu_{{\rm Ly}\alpha}}

\newcommand{\HI}{H{\sc ~i}}

\newcommand{\HeI}{He{\sc ~i}}
\newcommand{\HeII}{He{\sc ~ii}}

\begin{document}

\title{Lyman-$\alpha$ transfer in primordial hydrogen recombination}

\author{Christopher M. Hirata}
\email{chirata@tapir.caltech.edu}
\affiliation{Caltech M/C 350-17, Pasadena, California 91125, USA}

\author{John Forbes}
\email{forbes@caltech.edu}
\affiliation{Caltech MSC 272, Pasadena, California 91126, USA}

\date{June 15, 2009}

\begin{abstract}
Cosmological constraints from the cosmic microwave background (CMB) anisotropies rely on accurate theoretical calculations of the cosmic recombination 
history.  Recent work has emphasized the importance of radiative transfer calculations due to the high optical depth in the \HI\ Lyman lines.  
Transfer in the Ly$\alpha$ line is dominated by true emission and absorption, Hubble expansion, and resonant scattering.  Resonant scattering causes 
photons to diffuse in frequency due to random kicks from the thermal velocities of hydrogen atoms, and also to drift toward lower frequencies due to 
energy loss via atomic recoil.  Past analyses of Ly$\alpha$ transfer during the recombination era have either considered a subset of these processes, 
ignored time dependence, or incorrectly assumed identical emission and absorption profiles.  We present here a fully time-dependent radiative transfer 
calculation of the Ly$\alpha$ line including all of these processes, and compare it to previous results that ignored the resonant scattering.  
We find a faster recombination due to recoil enhancement of the Ly$\alpha$ escape rate, leading to a reduction in the free electron density of 0.45\% at 
$z=900$.  This results in an increase in the small-scale CMB power spectrum that is negligible for the current data but will be a 0.9$\sigma$ 
correction for {\slshape Planck}.  We discuss the reasons why we find a smaller correction than some other recent computations.
\end{abstract}


\maketitle

\section{Introduction}

The cosmic microwave background (CMB) anisotropy has proven to be one of the most useful and robust cosmological probes.  The most recent example has 
been the {\slshape Wilkinson Microwave Anisotropy Probe} (WMAP) satellite, which has provided key information on the composition of the high-redshift 
Universe, the distance to the surface of last scattering (via the acoustic peak position), cosmic reionization, and the primordial power spectrum 
\cite{2007ApJS..170..377S, 2009ApJS..180..306D, 2009ApJS..180..330K}.  The development of precision cosmology with the CMB will continue with the 
upcoming launch of the {\slshape Planck} satellite.  {\slshape Planck} data will be essential for dark energy studies: its measurement of $\Omega_mh^2$ 
and $D_A(z=1100)$ will be needed to break degeneracies with $w(z)$ in supernova data and set the length of the baryon acoustic oscillation standard 
ruler, and the primordial power spectrum will be needed to interpret tests of the growth of structure using weak lensing and 
clusters \cite{2002PhRvD..65b3003H, 
2006astro.ph..9591A, 2009arXiv0901.0721A}.  The measurement of the primordial spectral index $n_s$ will be essential for constraining models of 
inflation; indeed it is the 
precision measurement of $n_s$ from WMAP (and upper limits on tensors) that have opened the era of tests of inflaton potentials 
\cite{2003ApJS..148..213P, 2009ApJS..180..330K}.

In addition to its versatility, one of the advantages of the CMB as a cosmological probe is that the generation of CMB anisotropies is understood from 
first principles: they arise due to linear perturbations on a homogeneous, isotropic background of baryons, photons, dark matter, and neutrinos.  
There are now several public codes that solve these equations and compute the angular power spectrum $C_\ell$ \cite{1996ApJ...469..437S, 
2000ApJ...538..473L} and whose numerical accuracy has now been checked to 1 part in 10$^3$ \cite{2003PhRvD..68h3507S}.  However the photons and baryons 
interact via Thomson scattering, so these codes require as input the ionization fraction $x_e(z)$.  This is a complicated non-equilibrium physics 
problem as one must track the level populations of recombining hydrogen and helium atoms, as well as following the effects of the emitted radiation and 
its re-absorption.

The first attempts to solve primordial recombination were carried out by Peebles \cite{1968ApJ...153....1P} and Zel'dovich et al 
\cite{1968ZhETF..55..278Z}, and the progress of CMB experiments has driven theorists to construct ever more accurate models of recombination.  Most CMB 
codes in common use today obtain $x_e(z)$ from the {\sc Recfast} code by Seager et al \cite{1999ApJ...523L...1S,2000ApJS..128..407S}, which is based on 
the multi-level atom (MLA) approximation.  The MLA follows the level populations of each type of atom considered (\HI, \HeI, \HeII) under the influence 
of bound-bound and bound-free transitions.  The optically thick lines (the \HI\ Lyman series and its \HeI\ and \HeII\ analogues) are treated using the 
Sobolev approximation \cite{1960mes..book.....S} to avoid the need to track the radiation field explicitly.  Two-photon decays from the \HI\ $2s$, 
\HeI\ $2^1S_0$, and \HeII\ $2s$ levels and their inverse process (two-photon absorption) are important and also included.

The picture that emerged from this work is one in which hydrogen recombination occurs far more slowly than the Saha equation predicts.  The reason is 
the production of radiation: a hydrogen atom that recombines directly to the ground level ($1s$) emits a Lyman continuum photon.  Once $\sim 1$ out of 
every 10$^8$ H atoms has recombined the Universe becomes optically thick to this radiation and each direct recombination is immediately followed by an 
ionization.  Recombination can therefore occur only indirectly, via recombination to the \HI\ excited levels ($nl$, $n\ge 2$) and radiative cascade to 
$1s$.  But here another bottleneck occurs: the buildup of radiation in the Ly$\alpha$ ($2p\rightarrow 1s$, 1216\AA) line.  In the absence of a sink for 
Ly$\alpha$ photons, a hydrogen atom in the $2p$ level can only reach the ground level by exciting another hydrogen atom to $2p$.  The intense thermal 
CMB radiation during the recombination epoch is easily capable of ionizing a hydrogen atom from the $n=2$ level, so again the process produces no net 
recombinations.  The two major processes that break the bottleneck are (i) the cosmological redshifting of photons out of the Ly$\alpha$ line, and (ii) 
the $2s\rightarrow 1s$ two-photon decay (the photons have a continuous energy distribution and hence are not immediately re-absorbed in hydrogen 
lines).  Similar physics applies at earlier epochs to helium recombination, albeit with some additional complications due to the more complex energy 
level structure.

In the past several years, many authors have re-investigated the recombination problem with attention to small physical effects, in particular 
neglected pathways that might allow hydrogen atoms to reach the $1s$ level.  In particular, the {\slshape Planck} mission will require much better 
accuracy than was desired in 1999 when the original version of {\sc Recfast} was written.  The new effects considered have included two-photon decays 
from the $ns$ and $nd$ levels of \HI\ \cite{2005AstL...31..359D, 2007MNRAS.375.1441W, 2008A&A...480..629C, 2008PhRvD..78b3001H, 2008AstL...34..289K} 
and the $n^1S$ and $n^1D$ levels of \HeI\ \cite{2005AstL...31..359D, 2007MNRAS.375.1441W, 2008PhRvD..77h3007H}; \HI\ continuum absorption of \HeI\ 
584\AA\ line radiation \cite{2008PhRvD..77h3006S, 2007MNRAS.378L..39K, 2008A&A...485..377R}; stimulated two-photon decays and two-photon absorption 
\cite{2006A&A...446...39C, 2006AstL...32..795K, 2008PhRvD..78b3001H}; Raman scattering \cite{2008PhRvD..77h3007H, 2008PhRvD..78b3001H}; resolution of 
the \HI\ $l$-sublevels \cite{2007MNRAS.374.1310C}; and forbidden transitions in \HeI\ \cite{2007MNRAS.375.1441W, 2008PhRvD..77h3006S, 
2008PhRvD..77h3008S, 2008A&A...485..377R}.  In some cases the corrections to the $C_\ell$s were $>1$\%, sufficient to bias the $n_s$ measurement from 
{\slshape Planck} by several sigma \cite{2006MNRAS.373..561L}.  Some of these corrections are now incorporated into the most recent version of {\sc 
Recfast} \cite{2008MNRAS.386.1023W}.

An additional correction is the deviation from strict Sobolev behavior in the very optically thick \HI\ Ly$\alpha$ line ($2p\rightarrow 1s$).  This is 
the subject of this paper.  The Sobolev approximation is based on the assumptions of (i) identical absorption and emission profiles, (ii) complete 
frequency redistribution in each line scattering, (iii) absence of any other absorption or emission processes active in the same frequency range as the 
line, and (iv) quasi-stationarity, i.e. that the level populations and radiation field change little during the time it takes for a photon to redshift 
through the line.  None of these approximations are quite valid during the recombination era.  Previous authors have relaxed some of these assumptions 
individually, but there exists no comprehensive treatment.  Krolik \cite{1989ApJ...338..594K, 1990ApJ...353...21K} considered the partial 
redistribution during Lyman-$\alpha$ scattering, and found only a small correction due to atomic recoil.  Rybicki \& dell'Antonio 
\cite{1994ApJ...427..603R} relaxed the quasi-stationary assumption and found that the relaxation timescale for the Ly$\alpha$ line was a factor of a 
few shorter than the recombination time; in the context of modern experiments this would probably imply a significant correction to the recombination 
history. Hirata \& Switzer \cite{2008PhRvD..77h3007H} argued in the context of the \HeI\ 584\AA\ line that the deviation of absorption versus emission 
profiles would lead to a somewhat increased escape probability, an effect which was verified to be important for \HI\ Ly$\alpha$ by Hirata 
\cite{2008PhRvD..78b3001H}.  There are also recent treatments of diffusion and atomic recoil by Grachev \& Dubrovich \cite{2008AstL...34..439G} and 
quasi-stationarity by Chluba \& Sunyaev \cite{2008arXiv0810.1045C}.  However there is not yet a treatment relaxing all of (i--iv) simultaneously.

In this paper, we will first review the physics of the recombination era with an emphasis on the role of Ly$\alpha$ escape (\S\ref{sec:bk}).  Next we 
describe a numerical method for solving the Fokker-Planck equation in the vicinity of Ly$\alpha$ and grafting it on to an existing MLA code, and 
present the results obtained by this technique (\S\ref{sec:num}).  We then describe an analytic approach to the Ly$\alpha$ escape, which is much 
simpler than the fully numerical technique and captures the essential physics (\S\ref{sec:ana}). We describe implications for precision CMB 
observations in \S\ref{sec:cmb}.  We conclude in \S\ref{sec:conc}.  Appendix~\ref{app:sol} describes computation of two special functions $\chi(W,S)$ 
and ${\cal I}(W,S)$ introduced in this paper.

For ease of comparison, we use the same fiducial cosmology as in Ref.~\cite{2008PhRvD..78b3001H}: $\Omega_mh^2=0.13$, $\Omega_bh^2=0.022$, $T_{\rm 
CMB}=2.728\,$K, helium mass fraction $Y=0.24$, and an effective number of massless neutrinos $N_\nu=3.04$.  We draw heavily on the work of Hirata 
\cite{2008PhRvD..78b3001H}, where our MLA code was first presented.

\section{Background}
\label{sec:bk}

In this section, we begin with a review of the basic definitions (\S\ref{ss:def}).  We then review the MLA method and its extension to two-photon 
transitions (\S\ref{ss:mla}).  We then describe the physical formulation of the Ly$\alpha$ diffusion problem (\S\ref{ss:dif}).

\subsection{Definitions}
\label{ss:def}

Since this paper extends the code of Hirata \cite{2008PhRvD..78b3001H}, we begin with the same notation and basic equations as in that paper, and add 
new variables as needed for the Ly$\alpha$ problem.  The total density of hydrogen nuclei is $\nH\propto a^{-3}$ where $a$ is the scale factor.  
Ionization fractions are given by the free electron abundance $x_e\equiv n_e/\nH$ and free proton abundance $x_p\equiv n_p/\nH$ relative to hydrogen.  
Since this paper concerns the epoch after completion of helium recombination and before any hydrogen converts to molecular or intermediate (H$^-$, 
H$_2^+$, etc.) forms, we have $x_e=x_p$.  Photons are described by the phase space density $f(E)$.  The abundance of specific energy levels of the 
hydrogen atom will be given by $x_{nl}=n[$\HI$(nl)]/\nH$.  Level degeneracies $g_{nl}=2(2l+1)$, energies $E_{nl}=-h{\cal R}/n^2$, and Einstein 
coefficients $A_{ij}$ have their usual meaning; we denote the hydrogenic Rydberg (in frequency units) by ${\cal R}$, and set Boltzmann's constant equal 
to 1 so that temperature has units of energy.

\subsection{Multi-level atom and two-photon transitions}
\label{ss:mla}

The Hirata \cite{2008PhRvD..78b3001H} code follows bound-bound and bound-free transitions involving each level of \HI.  The treatment of bound-free 
transitions and of the matter temperature is not altered by this paper and the corresponding equations will not be repeated.  For the bound-bound case, 
the rate equation is
\begin{eqnarray}
\dot x_i|_{\rm bb} &=& \sum_{j>i} A_{ji}P_{ji}\left[(1+f_{ji+})x_j-\frac{g_j}{g_i}f_{ji+}x_i\right]
\nonumber \\ && + \sum_{j<i}
A_{ij}P_{ij}\left[\frac{g_i}{g_j}f_{ij+}x_j-(1+f_{ij+})x_i\right],
\label{eq:xibb}
\end{eqnarray}
where the sums are over levels $j$ that are above ($j>i$) or below ($j<i$) the energy of level $i$, and $f_{ji+}$ is the phase space density on the 
blue side of the line connecting levels $i$ and $j$.  The escape 
probability $P_{ji}$ is expressed in 
terms of the Sobolev optical depth $\tau_{ji}$ via
\begin{equation}
P_{ji} = \frac{1-e^{-\tau_{ji}}}{\tau_{ji}},
\label{eq:prob}
\end{equation}
and the optical depth is
\begin{equation}
\tau_{ji} = \frac{c^3n_{\rm H}}{8\pi H\nu_{ji}^3}
A_{ji}\left(\frac{g_j}{g_i}x_i-x_j\right).
\end{equation}
Of particular importance to us is the optical depth in the Ly$\alpha$ ($2p\rightarrow 1s$) line, which is typically of order $10^8$--10$^9$.

The phase space density $f_{ji+}$ on the blue side of each line is required since these photons redshift into the line and can be absorbed or cause 
stimulated emission.  In most cases it can be treated as a blackbody, but in the case of the Lyman series lines one must consider the nonthermal nature 
of the radiation field.  In particular, photons emitted in the Ly$\beta$ line (1026\AA) will eventually redshift into Ly$\alpha$ and begin exciting 
atoms.  This ``feedback'' process \cite{2007A&A...475..109C, 2008PhRvD..77h3006S} is taken into account by looking up the phase space density that 
emerged from Ly$\beta$ at a previous epoch and taking this as input for the Ly$\alpha$ calculation.

Two-photon decays of the form
\begin{equation}
{\rm H}(nl) \rightarrow {\rm H}(1s) + \gamma + \gamma
\label{rx:hdecay}
\end{equation}
produce photons at a rate per unit frequency per H atom of
\begin{equation}
\Delta_{nl}(\nu) = \frac{d\Lambda_{nl}}{d\nu} \left[ (1+f_\nu)(1+f_{\nu'}) x_{nl} - \frac{g_{nl}}{g_{1s}} f_\nu f_{\nu'} x_{1s} \right],
\label{eq:dnl}
\end{equation}
where $d\Lambda_{nl}/d\nu$ is the spontaneous two-photon decay rate.  One can easily incorporate the total decay rate
\begin{equation}
\dot x_{nl}|_{2\gamma} = -\int_{(1-n^{-2}){\cal R}/2}^{(1-n^{-2}){\cal R}} \Delta_{nl}(\nu)\,d\nu
\label{eq:xnl}
\end{equation}
in the system of ODEs for the excited levels.  However one must also take account of the radiation produced: if one of the emitted photons has an 
energy exceeding Ly$\alpha$, it will eventually redshift into the Ly$\alpha$ resonance and excite a hydrogen atom.  This is done by the virtual level 
method \cite{2008PhRvD..78b3001H}, which implements Eq.~(\ref{rx:hdecay}) by introducing {\em purely as a mathematical device} a new virtual level of 
the hydrogen atom with energy $E_{1s}+h\nu$ (where $\nu$ is the frequency of the higher-energy photon) and infinitesimal degeneracy.  The decay in 
Eq.~(\ref{rx:hdecay}) can then be treated as a sequence of 
one-photon decays.  The required choice of effective one-photon rate coefficients required for this procedure to work are given in \S IVA of 
Ref.~\cite{2008PhRvD..78b3001H}; these choices also account for two-photon absorption.  The same procedure also applies to other two-photon 
processes such as Raman scattering and two-photon recombination/ionization.  Most importantly, the same feedback machinery used for one-photon 
transitions is automatically applied to the two-photon transitions.

The two-photon differential decay rate $d\Lambda_{nl}/d\nu$ contains resonances associated with the allowed 1-photon decays.  For example, the 
two-photon decay rate from $3d\rightarrow 1s$ has a resonance corresponding to the sequence
\begin{equation}
{\rm H}(3d) \rightarrow {\rm H}(2p) + \gamma({\rm H}\alpha),
\;\;\;
{\rm H}(2p) \rightarrow {\rm H}(1s) + \gamma({\rm Ly}\alpha).
\label{rx:seq}
\end{equation}
These resonances are physical since such decays actually do produce two photons \cite{2008A&A...480..629C}.  They do however 
result in very large decay rates if one naively computes the quantity
\begin{equation}
\Lambda^{\rm tot}_{3d} = \frac12\int_0^{8{\cal R}/9} \frac{d\Lambda_{3d}}{d\nu} d\nu = 6.5\times 10^7 \,{\rm s}^{-1}
\end{equation}
(compare to 8.2$\,$s$^{-1}$ for the $2s$ level).  Naively the existence of such a rapid $3d\rightarrow 1s$ two-photon decay process would dramatically 
accelerate recombination, but since most of these decays produce photons within the Ly$\alpha$ line this turns out not to be the case: almost every 
$3d\rightarrow 1s$ two-photon decay is immediately undone by Ly$\alpha$ absorption.  Ultimately the inclusion of these two-photon decays produces only 
${\cal O}(1$\%$)$ corrections to the recombination history when one tracks the radiation field as well \cite{2008PhRvD..78b3001H}.  It is important to 
note 
that this result can only be obtained by consideration of radiative transfer in the Ly$\alpha$ line \cite{2007MNRAS.375.1441W, 2008A&A...480..629C}.

Overall, this yields a system of ordinary differential equations (ODEs) for the atomic level populations.  These have explicit dependence on their 
history so the results of the integration must be stored for future reference.  Time steps are equally spaced in $\ln a$, with a fiducial choice of 
$\Delta\ln a = 4.25\times 10^{-5}$.  The excited atomic levels are treated using the steady-state approximation, i.e. assuming that the atom reaches 
the ground state or is ionized in a time short compared to the recombination timescale.

\subsection{The Ly$\alpha$ diffusion problem}
\label{ss:dif}

The treatment of radiative transfer in the Lyman series described in \S\ref{ss:mla} correctly describes many processes.  It contains correct emission 
and absorption profiles for all of the Lyman lines, including interference between neighboring resonances, and is fully time-dependent.  However, one 
key piece of physics is missing: it neglects the Doppler shift due to the motion of the atoms.  For continuum processes far from resonance this is a 
minor error.  It is a major omission in the vicinity of the Lyman lines, where repeated scattering of photons off of hydrogen atoms can cause them 
to undergo a random walk in frequency (frequency diffusion).  This is especially true for Ly$\alpha$ because of its very high scattering-to-absorption 
ratio.  This section is devoted to a qualitative discussion of the physical processes involved; quantitative computations are deferred to 
\S\ref{sec:num}.

\begin{figure}
\includegraphics[width=3.2in]{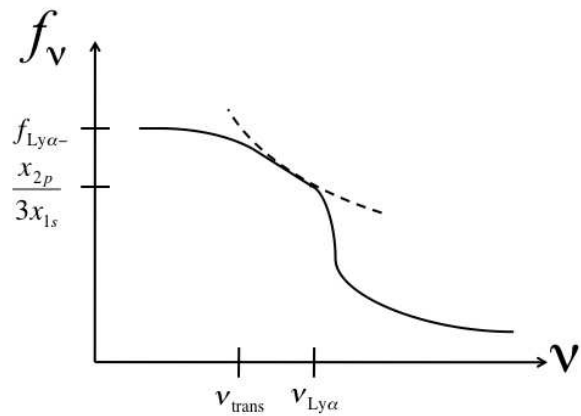}
\caption{\label{fig:pic1}A schematic representation of the photon phase space density near Ly$\alpha$.  The dashed line shows the chemical equilibrium 
solution, Eq.~(\ref{eq:chem}).  The true solution (solid line) comes to equilibrium near line center.  As one moves to the red side of the line, the 
rate of absorption, emission, and scattering decrease, and at some point ($\nu\sim\nu_{\rm trans}$) the photons fall out of chemical equilibrium and 
redshift out of the line.}
\end{figure}

The behavior of the radiation intensity near the Ly$\alpha$ line is a competition among several effects.  The resonant scattering of photons off of 
hydrogen atoms,
\begin{equation}
{\rm H}(1s) + \gamma \rightarrow {\rm H}(2p)_{\rm virtual} \rightarrow {\rm H}(1s) + \gamma,
\label{rx:scat}
\end{equation}
allows photons to exchange energy with the kinetic degrees of freedom of the matter but does not change the number of photons in the Ly$\alpha$ 
resonance.  Therefore if only this process were active it would drive the radiation intensity toward a modified blackbody 
$f_\nu=[e^{(h\nu-\mu)/\Tm}-1]^{-1}$ where the chemical potential $\mu$ is a constant.  Since near Ly$\alpha$ $f\ll 1$ the Bose-Einstein statistics of 
the photon are negligible and we expect $f_\nu\propto e^{-h\nu/\Tm}$.  A second process is the Hubble expansion, which moves photons to lower 
frequency at a constant rate $\dot\nu=-H\nu$.  A third process is the ``true'' emission and absorption of the Ly$\alpha$ photons -- that is, 
emissions and absorptions that are not part of the scattering process (Eq.~\ref{rx:scat}).  In this case, the line profile approaches the form shown in 
Figure~\ref{fig:pic1}: near line center the emission, absorption, and scattering are dominant.  The emission and absorption set the phase space density 
of photons 
at line center to be that of equilibrium, $f(\nula)=x_{2p}/(3x_{1s})$, and the scattering then gives the frequency dependence
\begin{equation}
f_\nu = \frac{x_{2p}}{3x_{1s}}e^{-h(\nu-\nula)/\Tm}.
\label{eq:chem}
\end{equation}
At some frequency $\nu_{\rm trans}$ sufficiently far to the red side of the Ly$\alpha$ line, a transition occurs where the time to redshift out of the 
line becomes 
shorter than the time to diffuse back to line center.  Beyond this point, the phase space density approaches a constant $f_{{\rm Ly}\alpha-}$, which is 
related to the net flux of photons to the red side of the Ly$\alpha$ line via
\begin{equation}
\frac{d(\#\,\rm photons)}{dV\,dt} = \frac{8\pi H}{\lambda_{{\rm Ly}\alpha}^3} f_{{\rm Ly}\alpha-}.
\label{eq:flux}
\end{equation}
(Here $8\pi H/\lambda_{{\rm Ly}\alpha}^3$ is simply the number of photon modes per unit volume per unit time that redshift through the frequency 
$\nula-\epsilon$.)  Qualitatively, the rate at which photons redshift out of Ly$\alpha$
is determined by Eq.~(\ref{eq:chem}) at the transition frequency, combined with Eq.~(\ref{eq:flux}).  This in turn gives the net $2p\rightarrow 1s$ 
decay rate (after a correction involving $f_{{\rm Ly}\alpha+}$ for photons that redshift {\em in} to Ly$\alpha$ and excite atoms is applied).  In 
equations, we have
\begin{equation}
\dot x_{2p\rightarrow 1s} = \frac{8\pi H}{\nH\lambda_{{\rm Ly}\alpha}^3} \left[\frac{x_{2p}}{3x_{1s}}e^{-h(\nu_{\rm trans}-\nula)/\Tm} - f_{{\rm 
Ly}\alpha+} \right].
\label{eq:peebles}
\end{equation}
This equation with $\nu_{\rm trans}\approx\nula$ and $f_{{\rm Ly}\alpha+}=(e^{h\nula/\Tr}-1)^{-1}$ gives the Peebles \cite{1968ApJ...153....1P} 
transition rate, equivalent to the Sobolev rate in the limit of $\tau\gg 1$; an excellent description of the physics can be found in \S6 of the book by 
Peebles \cite{1993ppc..book.....P}.  The same equation, with various estimates of the ``effective'' $\nu_{\rm trans}$, is the underlying conceptual 
reason for the accelerated recombination found by authors who considered repeated Ly$\alpha$ scattering \cite{2008AstL...34..439G}.  (Recombination is accelerated rather than decelerated because $\nu_{\rm trans}<\nula$.)

The role of emission and absorption in the Ly$\alpha$ damping wings must also be considered.  It was at first argued that because the emission and 
absorption profiles are the same (they are both Voigt profiles), this process tends to smooth out the frequency dependence of $f_\nu$ and suppress the 
boost factor $e^{-h(\nu_{\rm trans}-\nula)/\Tm}$ in Eq.~(\ref{eq:peebles}) \cite{1990ApJ...353...21K}.  However, true absorption of
Ly$\alpha$ photons is actually a multiple-photon process: the virtual H($2p$) atom must not decay back to the ground state, but rather absorb another 
photon, e.g.
\begin{equation}
{\rm H}(1s) + \gamma({\rm Ly}\alpha) + \gamma({\rm H}\alpha) \rightarrow {\rm H}(3s,3d).
\label{rx:3sd}
\end{equation}
It follows that the Ly$\alpha$ absorption profile actually depends on the color temperature of the ambient CMB near H$\alpha$, because of energy 
conservation: if the first photon is from the red tail of Ly$\alpha$ then the second must be from the blue tail of H$\alpha$ and vice versa.  In 
contrast, the emission profile, i.e. the reverse of Eq.~(\ref{rx:3sd}), is simply the Voigt profile (with small corrections due to neighboring 
resonances and stimulated emission).  Thus in the cosmological context, the absorption profile is enhanced relative to the emission profile in the blue
wing, and suppressed in the red wing, so true absorption and emission also tend to establish a red tilt to the radiation spectrum.
We can also understand this effect
thermodynamically: since true absorption of a Ly$\alpha$ photon followed by true emission (a process termed ``incoherent scattering'' by Krolik) 
exchanges the Ly$\alpha$ photon's energy with the low-energy H$\alpha$ photons, it follows that this process also tends to drive the radiation spectrum 
toward a modified blackbody, but this time with temperature $\Tr$.  At high redshifts we have $\Tr\approx\Tm$ and so once again Eq.~(\ref{eq:chem}) 
should apply if true emission and absorption are dominant processes.  Their consideration leads to a spectrum similar to that of Fig.~\ref{fig:pic1} 
and hence to a transition frequency and enhanced redshifting rate just as does Ly$\alpha$ scattering.  This effect has been seen in the consideration 
of two-photon decays; see e.g. \S VB of Ref.~\cite{2008PhRvD..78b3001H}.

A fourth process is the $2s\rightarrow 1s$ two-photon continuum: a small fraction of this continuum overlaps the red damping tail of Ly$\alpha$ 
(especially when modifications due to stimulated emission are considered) and this should be taken into account.

A complete theory of recombination must take full account of all these effects, and yield predictions for the $2p\rightarrow 1s$ decay rate and 
Ly$\alpha$ radiation profile.  It should go well beyond the conceptual discussion of ``$\nu_{\rm trans}$'' and account for non-time-steady effects.

The basic strategy in both our numerical and analytical methods is to excise the region immediately surrounding the Ly$\alpha$ line from the 
aforementioned two-photon radiative transfer code and replace it with a method that solves the diffusion equation at high resolution.  A hybrid method 
is necessary because the resolution required near Ly$\alpha$ (we must resolve the Doppler width of the line) would be too expensive if applied to the 
entire spectrum.

\section{Numerical method}
\label{sec:num}

In our numerical method, we treat the photons near the Ly${\alpha}$ using a time-dependent Fokker-Planck method.  This replaces the complicated 
redistribution of photons with a partial differential equation (PDE) that can be solved numerically.  We first set up this equation and its method of 
solution, and then comment on the initial and boundary conditions and interface to the MLA code.  The Fokker-Planck and MLA codes are interdependent, 
since the Fokker-Planck code requires boundary conditions and level populations from the MLA code, whereas the MLA code needs the Fokker-Planck code to 
determine the Ly$\alpha$ decay rate and the flux of photons emerging from the red wing of Ly$\alpha$.  We solve this problem by alternately running 
each code until convergence is achieved.  An interface script passes key variables between the two codes, such as the phase space density of photons 
that redshift across the frequency boundaries.

We divide the region of frequency near the Ly$\alpha$ into bins spaced equally in $\ln\nu$, usually with spacing $\Delta\ln\nu=8.5\times 10^{-6}$.  
This is fine enough that it takes many steps for a photon to redshift through a Doppler width (the condition is that $\Delta\ln\nu\ll \sqrt{\Tm/\mH 
c^2}$).  We denote by $N_i$ the number of photons per H nucleus per frequency bin in the $i^{\rm th}$ bin, and the frequency associated with the 
$i^{\rm th}$ bin by $\nu_i$.  This is related to the phase space density via
\begin{equation}
N_i = \frac{8\pi\nu_i^3\Delta\ln\nu}{c^3\nH}f(\nu_i).
\label{eq:Ni}
\end{equation}
This evolves with time according to Hubble redshifting, emission, absorption, and scattering:
\begin{equation}
\dot N_i = \dot N_i|_{\rm H} + \dot N_i|_{\rm em} + \dot N_i|_{\rm ab} + \dot N_i|_{\rm sc}.
\label{eq:boltz-ni}
\end{equation}

We use $M=2001$ bins in our standard case, and place the $i_0{^{\rm th}}$ bin [where $i_0=\frac12(M-1)=1000$] at Ly$\alpha$ line center.  Then
\begin{equation}
\nu_i = \nula\exp [(i-i_0)\Delta\ln\nu].
\end{equation}

\subsection{Transport in the Ly$\alpha$ line}
\label{ss:trans}

Here we describe the computation of each term in Eq.~(\ref{eq:boltz-ni}).

\subsubsection{Emission}

We next consider the true emission of Ly$\alpha$ photons.  In the case of emission, the photons arise via two-photon decays from the 
$n\ge 3$ states, Eq.~(\ref{rx:seq}).  In the limit where we are close enough to Ly$\alpha$ line center to neglect the other resonances and the 
variation in photon phase space factors across the line, we may write this as
\begin{equation}
\dot N_i|^{\rm R}_{\rm em} = \sum_u x_u \Gamma_{u\rightarrow 2p} P_{2p \rightarrow 1s} \phi_{\rm V}(\nu_i) \Delta\nu_i.
\label{eq:emission1}
\end{equation}
This is a sum over all higher-energy states of hydrogen $u$, where
$\Gamma_{u\rightarrow 2p}$ is the rate per second at which those states decay to the 2p state, $\phi_{\rm V}(\nu)$ is the Voigt distribution,
$P_{2p \rightarrow 1s}$ is the branching fraction for $2p$ to decay to the ground level ($P_{2p \rightarrow 1s}=1$ in vacuum),
and $\Delta\nu_i$ is the bin width.

Rather than keep track of all the variables necessary to make that calculation, we retrieve the Ly$\alpha$ production
rate in photons per hydrogen atom per Hubble time from the interface (\S\ref{ss:extdata}).
Equation (\ref{eq:emission1}) simply reduces to
\begin{equation}
\dot N_i|^{\rm R}_{\rm em} = H \Pi \phi_{\rm V}(\nu_i) \Delta \nu_i
\label{eq:em-final1}
\end{equation}
where
\begin{equation}
\Pi = \frac1H\sum_u x_u \Gamma_{u\rightarrow 2p} P_{2p \rightarrow 1s}
\label{eq:Pi}
\end{equation}
denotes the Lyman $\alpha$ production rate.

In reality the neglect of photon phase space factors and other resonances in Eq.~(\ref{eq:em-final1}) is not correct to the desired accuracy.  We have 
thus implemented a ``corrected'' form of the equation,
\begin{equation}
\dot N_i|_{\rm em} = H \Pi {\cal E}(\nu_i) \phi_{\rm V}(\nu_i) \Delta \nu_i,
\label{eq:em-final}
\end{equation}
where ${\cal E}(\nu)$ is the correction function.  It should be given by
\begin{equation}
{\cal E}(\nu) = \frac{\sum_{nl} x_{nl} (d\Lambda_{nl}/d\nu)(1+f_{\nu'})}{\sum_{nl} x_{nl} (d\Lambda_{nl}^{\rm R}/d\nu)(1+f_{nl,2p})},
\label{eq:e-nu}
\end{equation}
where $d\Lambda_{nl}^{\rm R}/d\nu$ is the resonance profile approximation to $d\Lambda_{nl}/d\nu$, i.e.
\begin{equation}
\frac{d\Lambda_{nl}^{\rm R}}{d\nu} = \frac{512\alpha_{\rm fs}^6\nu_{nl,2p}^3}{19683(2l+1){\cal R}a_0^2}
\frac{\left|\langle nl||r||2p\rangle\right|^2}{(\nu-\nula)^2}.
\label{eq:lR}
\end{equation}
[This equation is derived from Eq.~(B5) of Ref.~\cite{2008PhRvD..78b3001H}, taking only the leading-order $(\nu-\nula)^{-2}$ term and recalling that 
$\langle 2p||r||1s\rangle=2^{15/2}a_0/3^{9/2}$ and $\nula=\frac34{\cal R}$.]  There should technically be stimulated emission factors of $1+f_\nu$ 
and $1+f_{{\rm Ly}\alpha}$ in Eq.~(\ref{eq:e-nu}), but in the vicinity of Ly$\alpha$ $f_\nu<10^{-11}$ during our period of integration so we leave 
these out.  Also ${\cal E}(\nu)$ should technically be computed in the rest frame of the hydrogen atom rather than in the comoving frame, but since 
${\cal E}(\nu)$ varies extremely slowly with frequency (it does not possess a resonance at Ly$\alpha$) this correction is unimportant.

Note that in Eq.~(\ref{eq:e-nu}), the $2s$ level should be included in 
the numerator since it is possible for a $2s\rightarrow 1s$ two-photon decay to produce emission within the frequency range of the Fokker-Planck
code.  In the blue wing of Ly$\alpha$ ($\nu>\nula$), one should replace this with the $2s\rightarrow 1s$ Raman scattering rate,
\begin{equation}
\frac{d\Lambda_{2s}}{d\nu}(1+f_{\nu'}) \rightarrow \frac{dK_{2s}}{d\nu}f_{\nu'}.
\end{equation}
[Note that there is no need to include $2s$ in the denominator, since the resonance approximation for its decay rate, Eq.~(\ref{eq:lR}), is zero.]
Technically one should also include continuum states in the sum, but since the true two-photon emission is dominated by decays from $n\ge 3$ states 
rather than direct decays from the continuum we will not include the latter here.

Equation (\ref{eq:e-nu}) is in general quite complicated.  It can be evaluated under the approximation of Boltzmann equilibrium of the low-lying 
excited states.  (The biggest exception to this rule is the $2s:2p$ ratio, which deviates from statistical equilibrium by 0.1\% at $z=1190$, 1\% at 
$z=950$, and 10\% at $z=790$.  The $3s:2p$, $3p:2s$, and $3d:2p$ ratios remain in Boltzmann equilibrium to $<1$\% throughout at all $z>700$.)  This 
allows us to construct a function ${\cal E}(\nu,T)$.  This can be split into two contributions
\begin{equation}
{\cal E}(\nu,T) = {\cal E}_{2s}(\nu,T)+{\cal E}_{n\ge3}(\nu,T)
\end{equation}
coming from the $2s$ and $n\ge 3$ levels respectively.  Note that at $\nu=\nula$ we must have ${\cal E}_{n\ge 3}\rightarrow 1$ and ${\cal 
E}_{2s}\rightarrow 0$.
The functions ${\cal E}(\nu,T)$ is fit to $<0.3$\% accuracy over the range $|\vartheta|<0.01$ and $T<4700\,$K, where 
$\vartheta=\nu/\nula-1$, by
\begin{equation}
{\cal E}_{n\ge3}(\nu,T) = e^{-5.4\vartheta}
\end{equation}
and
\begin{equation}
{\cal E}_{2s}(\nu,T) = 92.5e^{6.0\vartheta}
\frac{e^{h\nu_{{\rm H}\alpha}/T}|\vartheta|^3}{|e^{\vartheta h\nula/T}-1|}\frac1{1 + 0.321e^{-h\nu_{{\rm Pa}\alpha}/T}}.
\end{equation}
[The first fraction in this equation is physically motivated by the low-energy photon phase space factor $\propto|\vartheta|^3$, the thermal stimulated 
emission ($1+f$) or 
absorption ($f$) phase space density $1/|e^{\vartheta h\nula/T}-1|$, and the Boltzmann enhancement of $n=2$ relative to $n=3$ levels $e^{h\nu_{{\rm 
H}\alpha}/T}$.  The second fraction takes into account the fact that some of the Ly$\alpha$ emission is preceded by H$\beta$ emission from the $n=4$
levels, with $e^{-h\nu_{{\rm Pa}\alpha}/T}$ representing the Boltzmann suppression of $n=4$ relative to $n=3$ hydrogen atoms.]

The Voigt profile $\phi_{\rm V}(\nu)$ is computed using the integral formulation \cite{2007MNRAS.375.1043Z}, except in the far damping wings where we 
switch to the asymptotic expansion for $|\nu-\nula|\gg\sigma_\nu$.

\subsubsection{Absorption}

True absorption is the inverse process of true emission, so the same matrix element applies to both cases.  In particular, the ratio of 
two-photon absorption from $1s$ to a given energy level $u$ is related to the rate of emission via:
\begin{equation}
\frac{\dot N|_{\rm em}}{\dot N|_{\rm ab}} = -\frac{g_{1s}x_u(1+f_\nu)(1+f_{\nu'})}{g_ux_{1s}f_\nu f_{\nu'}},
\end{equation}
where $f_\nu$ and $f_{\nu'}$ are the phase space densities associated with the two photons.  We take $\nu$ to represent the frequency of the photon 
near Ly$\alpha$ and $\nu'$ to represent that of the low-frequency photon.  Since the lower-frequency photon comes from a blackbody distribution, we 
have
\begin{equation}
\frac{1+f_{\nu'}}{f_{\nu'}} = e^{h\nu'/\Tr}.
\end{equation}
Further assuming that the $u$ level is in Boltzmann equilibrium with $2p$ (a good approximation for the $n\le4$ levels), we have
\begin{equation}
x_u = \frac{g_u}{g_{2p}}e^{-(E_u-E_{2p})/\Tr},
\end{equation}
so
\begin{equation}
\frac{\dot N|_{\rm em}}{\dot N|_{\rm ab}} =-
\frac{x_{2p}}{3x_{1s}f_\nu} e^{(-E_u + E_{2p} + h\nu')/\Tr}.
\end{equation}
Using conservation of energy to find that $h\nu'=E_u-E_{1s}-h\nu$, we can simplify this to
\begin{equation}
\frac{\dot N|_{\rm em}}{\dot N|_{\rm ab}} =-
\frac{x_{2p}}{3x_{1s}f_\nu} e^{-h(\nu-\nula)/\Tr}.
\end{equation}
This ratio applies for all of the excited states $u$, so it must apply to the total true emission and absorption rates as well.  Thus we can solve for 
$\dot N_i|_{\rm ab}$:
\begin{equation}
\dot N_i|_{\rm ab} = -\frac{3x_{1s}f_\nu}{x_{2p}} e^{h(\nu_i-\nula)/\Tr}\dot N_i|_{\rm em}.
\label{eq:ebam1}
\end{equation}
We can simplify this further by defining the equilibrium number of photons per bin at line center,
\begin{equation}
N_{\rm eq} \equiv \frac{8\pi\Delta\ln\nu}{\nH\lambda_{{\rm Ly}\alpha}^3} \frac{x_{2p}}{3x_{1s}},
\label{eq:Neq}
\end{equation}
from which we convert Eq.~(\ref{eq:ebam1}) into
\begin{equation}
\dot N_i|_{\rm ab} = -\frac{N_i}{N_{\rm eq}} \left(\frac{\nula}{\nu_i}\right)^3 e^{h(\nu_i-\nula)/\Tr}\dot N_i|_{\rm em}.
\end{equation}
Using Eq.~(\ref{eq:em-final}), we arrive at
\begin{equation}
\dot N_i|_{\rm ab} = -\frac{H \Pi \phi(\nu_i){\cal E}(\nu_i) \Delta \nu_i}{N_{\rm eq}} \left(\frac{\nula}{\nu_i}\right)^3 e^{h(\nu_i-\nula)/\Tr} N_i.
\label{eq:ab-final}
\end{equation}
The value of $N_{\rm eq}$ is provided by the interface (\S\ref{ss:extdata}).

\subsubsection{Scattering}

We now consider the change in frequency of photons due to resonant scattering off of hydrogen atoms, Eq.~(\ref{rx:scat}). The typical fractional change 
in frequency is roughly $v_{\rm th}/c$ where $v_{\rm th}$ is the rms thermal velocity of the hydrogen atoms.  Nevertheless in a very optically thick 
line the net effect of many scatterings on the line profile may be important.  We therefore write the Ly$\alpha$ transport in terms of a Fokker-Planck 
operator \cite{1994ApJ...427..603R, 2004ApJ...602....1C, 2006MNRAS.367..259H, 2006ApJ...647..709R, 2008AstL...34..439G}.  In formulating such an 
operator, it is essential to be sure that the scattering term exactly conserves photons and preserves the equilibrium distribution $f_\nu\propto 
e^{-h\nu/\Tm}$ \cite{2006ApJ...647..709R}, even after discretization.

We define $F_i$ to represent the net flux of photons from bin $i+1$ to bin $i$ due to scattering.  In this way we have
\begin{equation}
\dot N_i|_{\rm sc} = F_i - F_{i-1}.
\end{equation}
This formulation guarantees exact conservation of photons even in the discretized problem.  In Fokker-Planck problems the fluxes are linear in the 
number of photons and its frequency derivative, i.e. in $N_i$ and $N_{i+1}-N_i$, so we write
\begin{equation}
F_i = -\zeta_i N_i + \eta_i N_{i+1},
\label{eq:fi}
\end{equation}
where $\zeta_i$ and $\eta_i$ are coefficients to be determined.
In the equilibrium modified blackbody distribution, and for logarithmically spaced bins $\Delta\nu_i\propto\nu_i$, we have
\begin{equation}
N_i \propto \nu_i^3 e^{-h\nu_i/\Tm},
\end{equation}
so in order for this to give zero net flux, we must have
\begin{equation}
\frac{\zeta_i}{\eta_i} = \frac{\nu_{i+1}^3}{\nu_i^3}e^{-h(\nu_{i+1}-\nu_i)/\Tm}.
\label{eq:ze}
\end{equation}

Equation (\ref{eq:ze}) provides one constraint for two free parameters $\zeta_i$ and $\eta_i$.  The other constraint must come from fixing the 
diffusion coefficient ${\cal D}(\nu)$ to the correct value.  We see that
\begin{equation}
\dot N_i|_{\rm sc} = \zeta_{i-1}N_{i-1} - (\zeta_i+\eta_{i-1}) N_i + \eta_i N_{i+1};
\end{equation}
Taylor-expanding $N_i$ allows us to write
\begin{equation}
\dot N_i|_{\rm sc} = \frac12(\eta_i+\zeta_{i-1})\frac{\partial^2N_i}{\partial i^2}
+ {\rm 0th,~1st~derivatives},
\end{equation}
so the diffusion coefficient is $\frac12(\eta_i+\zeta_{i-1})$ bin$^2\,$s$^{-1}$.  This can be written in the usual units of Hz$^2\,$s$^{-1}$ by 
multiplying by the square of the bin width,
\begin{equation}
{\cal D}(\nu) = \frac12(\eta_i+\zeta_{i-1})\Delta\nu^2.
\label{eq:dnu}
\end{equation}
We then compare to the actual diffusion coefficient \cite{2006MNRAS.367..259H}
\begin{equation}
{\cal D}(\nu) = H\nula \sigma_\nu^{2}\tau_{{\rm Ly}\alpha}f_{\rm S}\phi_{\rm V}(\nu),
\end{equation}
where $\sigma_\nu^2=\nula^2\Tm/(\mH c^2)$ is the variance of the Doppler shift distribution due to motion of H atoms.  The fraction of Ly$\alpha$ 
absorptions that result simply in scattering (as opposed to true absorptions) is $f_{\rm S}$; it is close to unity throughout the calculation, but a 
correct value is provided by the interface.

Since $\zeta_i$ and $\eta_i$ are 
slowly varying functions, we may replace $\zeta_{i-1}\rightarrow\zeta_i$ in Eq.~(\ref{eq:dnu}) and get
\begin{equation}
\frac12(\eta_i+\zeta_i) = \frac{H\nula \sigma_\nu^{2}\tau_{{\rm Ly}\alpha}\phi_{\rm V}(\nu)}{\Delta\nu^2}
\end{equation}
or
\begin{equation}
\eta_i+\zeta_i =
\frac{H\nula \sigma_\nu^{2}\tau_{{\rm Ly}\alpha}[\phi_{\rm V}(\nu_i)+\phi_{\rm V}(\nu_{i+1})]}{(\nu_{i+1}-\nu_i)^2}.
\label{eq:ze2}
\end{equation}
This and Eq.~(\ref{eq:ze}) are sufficient to determine $\zeta_i$ and $\eta_i$.

We handle the boundary conditions by disallowing any diffusion flux at either the red or blue boundary: $F_{-1}=F_{M-1}=0$, where $M$ is the number of 
bins.

The line profile for scattering, $\phi_{\rm V}(\nu)$, is in principle modified by the existence of neighboring resonances such as Ly$\beta$.
However, comparison of the Voigt profile to the actual cross section for $1s\rightarrow 1s$ scattering shows errors of $<2$\% in the frequency
range of interest $|\vartheta|<0.01$.  Since the $1s\rightarrow 1s$ scattering makes only a $\le0.45$\% correction to the recombination history, we 
ignore the ``correction to the correction.''

\subsubsection{Hubble expansion and integration algorithm}

For logarithmically spaced bins in frequency, it is easy to compute the effect of the Hubble expansion: when the Universe expands by an amount 
$\Delta\ln a = \Delta\ln\nu$, all photons simply shift into the next lowest frequency bin.
Thus to account for the Hubble expansion, the contents of each bin are shifted down by one
frequency bin at each time step:
\begin{equation}
N_i = N_{i+1}({\rm previous}).
\label{eq:ni}
\end{equation}
Since there are only a finite number of frequency bins, the values of
the number density of photons that redshift into the highest-frequency bin must
be determined by some other means. This value is denoted $N_{\rm in}$, and is provided
by the interface (\S\ref{ss:extdata}).
This method of manually shifting photons to the left requires logarithmically spaced frequency bins and time steps. Moreover,
the resolution $\Delta\ln\nu$ is tied to the time step.
(Despite this restriction, this method has the advantage of avoiding spurious numerical diffusion, which would arise if the Hubble expansion term were 
simply written as a differential operator with a discretized derivative.)


Our method of solving Eq.~(\ref{eq:boltz-ni}) is thus to apply an implicit ODE solver (backward Euler) to the emission, absorption, and scattering 
terms in Eq.~(\ref{eq:boltz-ni}), evolve forward one time step, and shift the photons according to Eq.~(\ref{eq:ni}).  We repeat this basic operation 
until we reach the desired final redshift $z_{\rm final}$.

The abundance of photons in the highest frequency bin, $N_{n-1}$, is not specified by the above algorithm.  Physically it is determined by the phase 
space density of photons redshifting into the line, in accordance with Eq.~(\ref{eq:Ni}).  This depends on the two-photon radiative transfer 
calculation and is provided by the interface code.

The emission, absorption, and scattering terms in the above equation can be written as a matrix equation:
\begin{equation}
\dot N_i|_{\rm em+ab+sc} = C_{ij}N_j,
\end{equation}
where $C_{ij}$ is a tridiagonal matrix.  We step forward using a backward Euler method:
\begin{equation}
\frac{N_i(t+\Delta t) - N_i(t)}{\Delta t} = C_{ij}(t+\Delta t)N_j(t+\Delta t);
\label{eq:be}
\end{equation}
this is a first-order method but this is sufficient because of the extremely small time step.  Inspection of the emission, absorption, and scattering 
terms shows that $C_{ij}$ is tridiagonal and hence Eq.~(\ref{eq:be}) is a tridiagonal linear system for $\{N_i(t+\Delta t)\}$.  The $M\times M$ 
tridiagonal system can be solved by the usual ${\cal O}(M)$ complexity method of using the $i=0$ equation to eliminate $N_0(t+\Delta t)$, then using 
the $i=1$ equation to eliminate $N_1(t+\Delta t)$, and so on.

\subsection{Interface}
\label{ss:extdata}

The evolution equation for the $\{N_i\}$ is only part of the recombination problem; it must interface to the multi-level atom code with the proper 
boundary conditions.  This problem is considered here.  The basic approach is iterative: the multi-level atom code is run first, to generate a table of 
input data for the Fokker-Planck code.  Then the outputs of the Fokker-Planck code are used to apply corrections to the multi-level atom code, and so 
on until convergence is reached.

The data passed from the multi-level atom code to the Fokker-Planck code at each time step are:
\newcounter{MLA}
\begin{list}{\arabic{MLA}. }{\usecounter{MLA}}
\item The matter and radiation temperatures.
\item The $2p$ state width $\Gamma_{2p}$ (inverse lifetime including all processes that depopulate the state, including Ly$\alpha$ decay, and 
bound-bound and bound-free absorptions);
\item The true emission rate of Ly$\alpha$ photons $\Pi(a)$, computed using Eq.~(\ref{eq:Pi}).
\item The equilibrium abundance of Ly$\alpha$ photons $N_{\rm eq}/\Delta\ln\nu$, computed using Eq.~(\ref{eq:Neq}).
\item The abundance of photons redshifting into the Fokker-Planck grid region per H nucleus per Hubble time, i.e. $N_{M-1}/\Delta\ln\nu$ at 
$\nu_{M-1}$.
\item The fraction $f_{\rm inc}$ of Ly$\alpha$ absorptions that result in true absorption instead of scattering.  This is determined by the branching 
ratios for transitions out of the $2p$ level of hydrogen.  Note that $f_{\rm S}=1-f_{\rm inc}$.
\item The optical depth to scattering in Ly$\alpha$ photons, $\tau_{{\rm Ly}\alpha}f_{\rm S}$.
\end{list}

In order to complete the iteration cycle, the Fokker-Planck code must return the corrections to Ly$\alpha$ transport to the MLA code.  This is done in 
several steps.  First, we turn off the two-photon transitions in the MLA code involving frequencies between $\nu_0$ and $\nu_{M-1}$.  Then we correct 
the usual equations for the Ly$\alpha$ line with correction factors that cause it to produce the same outputs (net $2p\rightarrow 1s$ decay rate and 
photon phase space density at $\nu_0$) as the Fokker-Planck code.  Explicitly, the standard equation for the Ly$\alpha$ decay rate is
\begin{equation}
\dot x_{2p\rightarrow 1s,{\rm std}}(a) = 
\frac{8\pi H}{\nH\lambda_{{\rm Ly}\alpha}^3} \frac{x_{2p}}{3x_{1s}} - \frac{N_{\rm in}[(\nula/\nu_{M-1})a]}{\Delta\ln\nu},
\end{equation}
where the rate of incoming photons is measured at the frequency $\nu_{M-1}>\nula$ at an earlier time since these are the photons that will reach 
Ly$\alpha$ line center at scale factor $a$.  [Compare to Eq.~(\ref{eq:peebles}).]  The standard equation for the rate at which photons redshift out of 
Ly$\alpha$ is
\begin{equation}
f_{{\rm Ly}\alpha-,{\rm std}}(a) = \frac{x_{2p}}{3x_{1s}}.
\end{equation}
We replace these with the equations
\begin{equation}
\dot x_{2p\rightarrow 1s}(a) = \xi_1(a) \dot x_{2p\rightarrow 1s,{\rm std}}(a)
\end{equation}
and
\begin{equation}
f_{{\rm Ly}\alpha-}(a) = \xi_2(a) f_{{\rm Ly}\alpha-,{\rm std}}(a).
\end{equation}

The correction factors $\xi_1(a)$ and $\xi_2(a)$ are determined by the Fokker-Planck code as follows.  The net decay rate is
$\dot x_{2p\rightarrow 1s}(a) = \sum_{i=0}^{M-1} \dot N_i|_{\rm em+ab+sc}$.
This equation is unstable as written if $\dot N_i$ is determined by plugging $N_i$ into the evolution equations.  The stable method is to find the 
change $\Delta N$ in $\sum_{i=0}^{M-1}N_i$ before and after the em+ab+sc time step, and write
\begin{equation}
\dot x_{2p\rightarrow 1s}(a) = \frac {H\Delta N}{\Delta\ln\nu}.
\end{equation}
We can then find $\xi_1(a)=\dot x_{2p\rightarrow 1s}(a)/\dot x_{2p\rightarrow 1s,{\rm std}}(a)$.

The red wing radiation correction factor $\xi_2(a)$ can be obtained by examining the phase space density of radiation emerging from the red wing of the 
line.  In the MLA code with correction factor, this phase space density at $\nu_0$ will be
\begin{equation}
f\left(\nu_0,\frac{\nula}{\nu_0}a\right) = \xi_2(a)\frac{x_{2p}(a)}{3x_{1s}(a)}.
\end{equation}
The true phase space density is however known from the Fokker-Planck code, so one can solve for $\xi_2(a)$.

Because $\xi_1(a)$ and $\xi_2(a)$ are correction factors and are generally close to unity, we expect faster convergence by having the Fokker-Planck 
code return $\xi_1(a)$ and $\xi_2(a)$ than absolute decay rates and phase space densities, so this is what we do.

The frequency spacing and time step in the Fokker-Planck code are $\Delta\ln\nu=\Delta\ln a=8.5\times10^{-6}$, which is 5 times finer than the 
MLA code of Ref.~\cite{2008PhRvD..78b3001H} ($\Delta\ln a = 4.25\times 10^{-5}$).  Therefore the data provided by the MLA code are interpolated onto 
the 
finer grid required by the Fokker-Planck code.

We find that only
two iterations of alternately running the Fokker-Planck and MLA codes are necessary.  In our fiducial case, the first iteration leads
to changes $|\Delta x_e|/x_e$ of at most $8.5\times 10^{-3}$; the second iteration leads to a maximum change of $5.3\times 10^{-5}$;
and the third iteration $1.4\times 10^{-6}$.

As a test, we have run the Fokker-Planck code with the scattering term $\dot N_i|_{\rm sc}$ turned off, and found agreement with the previous 
MLA code of Ref.~\cite{2008PhRvD..78b3001H}, with a maximum error $|\Delta x_e/x_e|$ of $4\times 10^{-5}$ for $700<z<1600$.

\subsection{Results}

\begin{figure}
\includegraphics[angle=-90,width=3.3in]{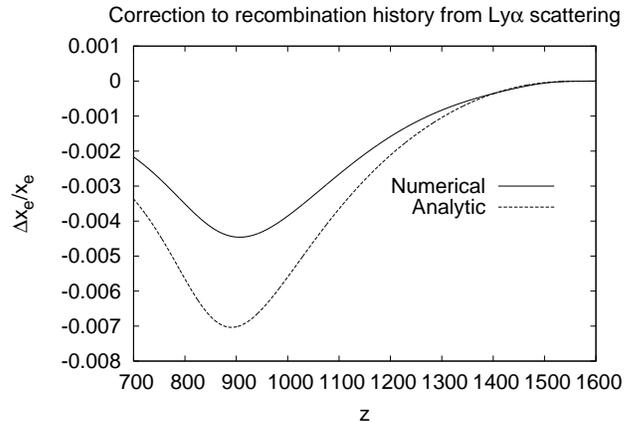}
\caption{\label{fig:dcorr}The correction to the recombination history due to Ly$\alpha$ diffusion.  The numerical computation of \S\ref{sec:num} is 
shown with a solid line, and \S\ref{sec:ana} with a dashed line.  Recombination is accelerated due to the additional redshifting of photons via atomic 
recoil.}
\end{figure}

The results from the Fokker-Planck code are shown in Fig.~\ref{fig:dcorr}.  This run began at $z_{\rm init}=1605.5$.  As expected from heuristic 
arguments (\S\ref{ss:dif}), the rate of recombination is accelerated by the inclusion of Ly$\alpha$ diffusion.

We have tested the convergence of our result with respect to the key numerical parameters.  For example, if we only include scattering
within $\pm500$ bins of the line center instead of the full $\pm1000$ bins, we find a maximum change in the ionization history $|\Delta x_e/x_e|$
of $10^{-5}$.  As an additional test, we tried using a 2.5$\times$ coarser frequency binning for the diffusion code (so that the diffusion 
code takes 2 instead of 5 time steps in each step of the MLA code).  The frequency range remained the same, so this corresponds to parameters 
$\Delta\ln\nu=\Delta\ln a=2.125\times 10^{-5}$ and $M=801$ bins.  This modification leads to a maximum change in the ionization history $|\Delta 
x_e/x_e|$ of $5\times 10^{-5}$.

\begin{figure}
\includegraphics[angle=-90,width=3.3in]{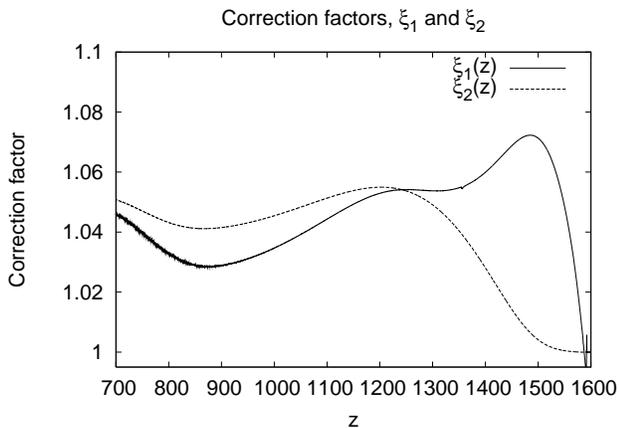}
\caption{\label{fig:xi}The correction factors $\xi_1(z)$ and $\xi_2(z)$.  The ``standard'' result for infinitesimally narrow Ly$\alpha$ line with
no damping wing or diffusion effects is $\xi_1(z)=\xi_2(z)=1$.
Note that over most of the recombination history these are greater than 1, implying a faster $2p\rightarrow 1s$ decay rate but also more photons
redshifting out of the Ly$\alpha$ line.  The latter effect will lead to more two-photon absorption at low redshifts.
[The glitch in $\xi_1(z)$ at $z\approx 1360$ is a startup transient from Ly$\beta$ at $z=z_{\rm init}$ redshifting into Ly$\alpha$; given that the 
overall effect of the scattering correction is $<0.6$\% in $C_\ell$, this is far too small to affect CMB results.]}
\end{figure}

The correction factors $\xi_1(z)$ and $\xi_2(z)$ are shown in Fig.~\ref{fig:xi}.

\begin{figure}
\includegraphics[angle=-90,width=3.3in]{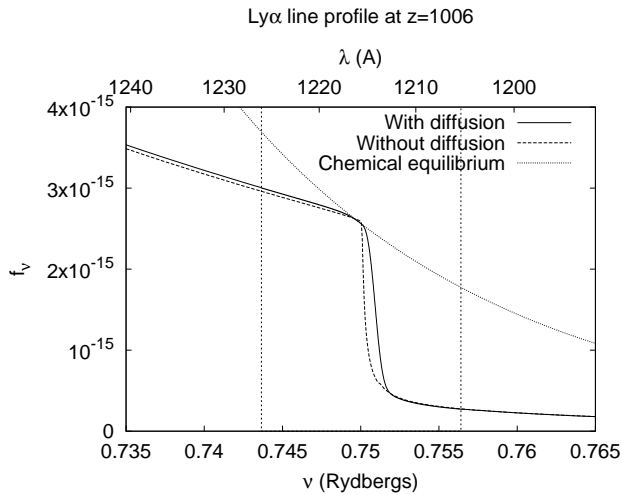}
\caption{\label{fig:lprof}The radiation spectrum in the vicinity of the Ly$\alpha$ line at $z=1006$.  The solid line shows the results of the 
diffusion code, and the dashed line shows the old code with no diffusion \cite{2008PhRvD..78b3001H}.  The vertical dotted lines show the 
boundaries of the Ly$\alpha$ diffusion region.  The ``chemical equilibrium'' curve shows the approximation $f_\nu\approx 
(x_{2p}/3x_{1s})e^{-h(\nu-\nula)/\Tr}$ that should be valid in the immediate vicinity of line center.  Note that $\nula=0.75$ Rydbergs.}
\end{figure}

We also show the radiation spectrum in Fig.~\ref{fig:lprof}.  The radiation phase space density outside the diffusion code boundary (i.e. $\nu<\nu_0$ 
or $\nu>\nu_{M-1}$) is obtained from the MLA code with virtual levels as in Ref.~\cite{2008PhRvD..78b3001H}, whereas for $\nu_0\le\nu\le\nu_{M-1}$ we 
have used the phase space density from the diffusion code.  We show the no-diffusion case (MLA + virtual levels only) with the dashed line.  The most 
noticeable effect of the diffusion is the increased intensity at $\nu>\nula$ due to Ly$\alpha$ photons diffusing to the blue side of the line.  There 
is also an enhancement in the number of photons redshifting out of the line, which directly affects the recombination rate.

\section{Analytic approximation}
\label{sec:ana}

We now consider a completely different approach to the Ly$\alpha$ diffusion problem, in which simple analytical calculations are used to estimate the 
correction to the Ly$\alpha$ escape rate.  This approach is a valuable complement to the fully numerical method: it contains additional approximations, 
but it provides a better understanding of the physics and a check of the much more sophistocated Fokker-Planck code/interface.  Analytic corrections 
(or more precisely, corrections based on reduction of the problem to a simple ODE) have been used by previous authors \cite{1989Ap.....30..211G, 
2008arXiv0810.1045C, 2008AstL...34..439G}.  The specific implementation here is an extension of the two-photon analysis of 
Ref.~\cite{2008PhRvD..78b3001H} to include Ly$\alpha$ diffusion.

In Ref.~\cite{2008PhRvD..78b3001H} the problem of emission and absorption in the Ly$\alpha$ damping wings was considered with no diffusion (i.e. 
without considering the change in frequency during a $1s\rightarrow 1s$ scattering).  The red and blue damping wings were handled separately since the 
radiative transfer phenomenology is different.  In both cases, the equation for the radiation field is written down and is approximated by its 
time-steady form.  Then:
\newcounter{wing}
\begin{list}{\arabic{wing}. }{\usecounter{wing}}
\item In the {\em red wing}, we find the correction to $f_\nu$, and based on the concept of the flux of photons [Eq.~(\ref{eq:flux})] 
we find a correction to the rate of Ly$\alpha$ escape.  Since the correction is always positive [Eq.~(\ref{eq:peebles})] the red wing corrections 
always increase the recombination rate, leading to lower ionization fraction.
\item
In the {\em blue wing}, we find the number $x_+(t)$ of 
spectral distortion photons in the blue wing of Ly$\alpha$ per H nucleus according to the time-steady equation.  This function starts at zero before 
recombination, reaches a positive maximum, and then declines to zero at late times as all photons redshift to lower frequencies.  An additional 
downward decay rate $\dot x_{2p\rightarrow 1s}=\dot x_+(t)$ is grafted on to the MLA code to account for the $2p\rightarrow 1s$ decays that are 
required early during recombination to build the spectral distortion, and then the excitations that occur later during recombination as this distortion 
redshifts into Ly$\alpha$.  Note that this process accelerates recombination at early times ($\dot x_+>0$) but delays it later ($\dot x_+<0$).
\end{list}

We now extend the treatment of Ref.~\cite{2008PhRvD..78b3001H} in the red (\S\ref{ss:red}) and blue (\S\ref{ss:blue}) wings.  In all cases we neglect 
the variation in phase space density (i.e. factors of $\nu/\nula$) across the Ly$\alpha$ line.

\subsection{Red wing}
\label{ss:red}

Without frequency diffusion, the radiative transfer equation, Eqs.~(82, 85, 87) of Ref.~\cite{2008PhRvD..78b3001H}, is
\begin{equation}
\frac{\dot f_\nu}{H\nu} =
\frac{\partial f_\nu}{\partial\nu} 
 -\frac{\bar W}{(\nu-\nula)^2}\left[
e^{h(\nu-\nula)/\Tr}f_\nu - \frac{x_{2p}}{3x_{1s}}
\right],
\label{eq:rte}
\end{equation}
where
\begin{equation}
\bar W = \frac{\tau_{{\rm Ly}\alpha}}{4\pi^2} \sum_{nl,n\ge3} \frac{2l+1}3 \frac{A_{nl,2p}}{e^{h\nu_{nl,2p}/\Tr} - 1}
\end{equation}
is the width over which Ly$\alpha$ is optically thick to true absorption.

We can add the frequency diffusion to this equation by recalling the diffusion term \cite{2006MNRAS.367..259H},
\begin{equation}
\dot f_\nu|_{\rm sc} = \frac\partial{\partial\nu} \left[ {\cal D}(\nu) \left(
\frac{\partial f_\nu}{\partial\nu} + \frac h{\Tm}f_\nu\right) \right].
\end{equation}
[Note that the $(h/\Tm)f_\nu$ term accounts for recoil.]
In the far damping wings, we have
\begin{equation}
{\cal D}(\nu) \approx \frac{H\nula\sigma_\nu^2\tau_{{\rm Ly}\alpha}f_{\rm S}A_{{\rm Ly}\alpha}}{4\pi^2(\nu-\nula)^2}
\end{equation}
(see Ref.~\cite{2006MNRAS.367..259H} and use the damping wing approximation to the Voigt profile for large $\nu-\nula$).
Adding this equation to Eq.~(\ref{eq:rte}) gives
\begin{eqnarray}
\frac{\dot f_\nu}{H\nu} &\approx&
\frac{\partial f_\nu}{\partial\nu}
 -\frac{\bar W}{(\nu-\nula)^2}\left[
e^{h(\nu-\nula)/\Tr}f_\nu - \frac{x_{2p}}{3x_{1s}}
\right]
\nonumber \\ &&
+ \frac{\sigma_\nu^2\tau_{{\rm Ly}\alpha}f_{\rm S}A_{{\rm Ly}\alpha}}{4\pi^2}
\nonumber \\ &&
\times
\frac\partial{\partial\nu} \left[ \frac1{(\nu-\nula)^2} \left(
\frac{\partial f_\nu}{\partial\nu} + \frac h{\Tm}f_\nu\right) \right].
\label{eq:rte2}
\end{eqnarray}
As in Ref.~\cite{2008PhRvD..78b3001H}, we make the change of variables
\begin{equation}
\nu = \nula + \frac{\Tr}h y
\end{equation}
and
\begin{equation}
f_\nu = \frac{x_{2p}}{3x_{1s}} \Phi(y).
\end{equation}
This, combined with dropping the $\dot f_\nu$ term (time-steady approximation) and taking $\Tm\approx\Tr$ (appropriate during the recombination era 
for the purposes of computing small corrections) simplifies Eq.~(\ref{eq:rte2}) to
\begin{equation}
0 = \frac{d\Phi}{dy} - \frac W{y^2}(e^y\Phi-1) + S\frac d{dy}\left[ y^{-2}\left(\frac{d\Phi}{dy} + \Phi\right) \right],
\label{eq:phi-red}
\end{equation}
where $W=h\bar W/\Tr$ and
\begin{equation}
S = \frac{\sigma_\nu^2\tau_{{\rm Ly}\alpha}f_{\rm S}A_{{\rm Ly}\alpha}h^3}{4\pi^2\Tr^3}.
\label{eq:S}
\end{equation}
This results in a dimensionless equation, Eq.~(\ref{eq:phi-red}), which depends on two constants $W$ and $S$.  The constant $W$ determines the strength 
of the true absorption: the Ly$\alpha$ line is optically thick to true absorption out to frequencies $\nula\pm W\Tr/h$.  The constant $S$ quantifies 
the importance of frequency diffusion relative to Hubble redshifting at frequencies $\nula\pm\Tr/h$.  In practice both are $\ll 1$.

\begin{figure}
\includegraphics[angle=-90,width=3.3in]{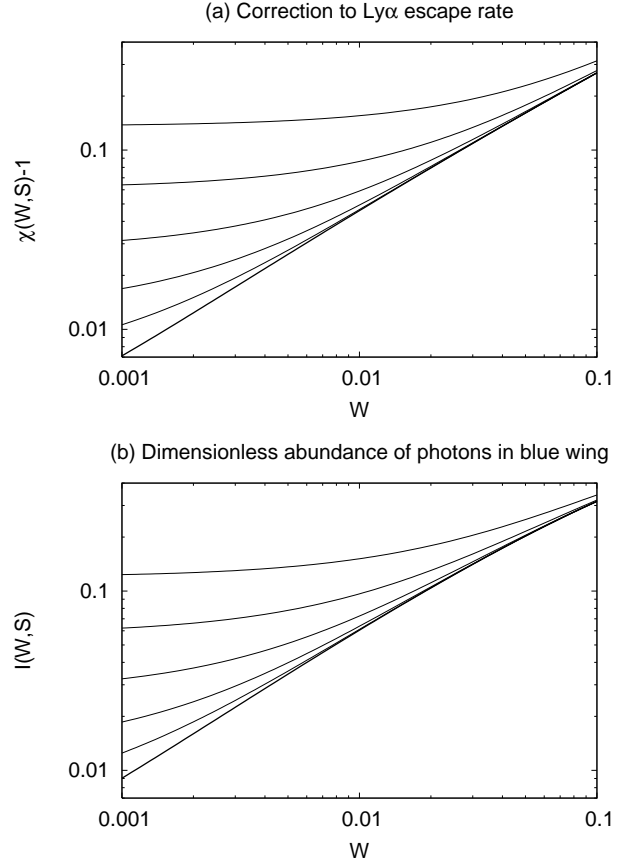}
\caption{\label{fig:chiI}The analytic correction factors $\chi(W,S)$ (top panel) and ${\cal I}(W,S)$ (bottom panel) associated with transport in the 
Ly$\alpha$ line.  The thick bold line shows the factors without frequency diffusion, i.e. for $S=0$.  The thin lines show the factors for $S=10^{-7}$, 
$10^{-6}$,
$10^{-5}$, $10^{-4}$, and $10^{-3}$ from bottom to top.  Note that both emission/absorption in the far damping wings (parameterized by $W$) and 
frequency diffusion (parameterized by $S$) tend to increase the escape rate and the number of distortion photons in the Ly$\alpha$ blue wing, as 
expected.}
\end{figure}

Our time-steady equation, Eq.~(\ref{eq:phi-red}), is very similar to Eq.~(93) of Ref.~\cite{2008PhRvD..78b3001H}, and it satisfies the same boundary 
condition: $\Phi=1$ at $y=0$, since at line center we reach equlibrium and have $f_{\nula}=x_{2p}/(3x_{1s})$.  The second-order 
differential operator complicates the solution and necessitates an additional boundary condition that 
$\Phi$ not diverge as $y\rightarrow -\infty$.  The numerical solution is presented in Appendix~\ref{app:sol}.  Just as in 
Ref.~\cite{2008PhRvD..78b3001H}, the correction to the net $2p\rightarrow 1s$ decay rate is
\begin{equation}
\Delta \dot x_\downarrow = \frac{A_{{\rm Ly}\alpha}}{\tau_{{\rm Ly}\alpha}} x_{2p}(\chi-1),
\label{eq:corr-red}
\end{equation}
where $\chi\equiv \Phi(y=-\infty)$.  The correction $\chi-1$ is now a function of the two dimensionless constants, $W$ and $S$.  It is shown 
graphically in Fig.~\ref{fig:chiI}.

\subsection{Blue wing}
\label{ss:blue}

The frequency diffusion in the blue damping wing leads to a modification of the number of spectral distortion photons $x_+(t)$ per H atom in the blue 
wing of Ly$\alpha$.  As in Ref.~\cite{2008PhRvD..78b3001H}, this can be approximated as
\begin{equation}
x_+ \approx \frac{8\pi\nula^2\Tr}{c^3\nH h}\int_0^\infty \Delta f_\nu\,dy,
\end{equation}
where $y$ is dimensionless frequency and $\Delta f_\nu$ is the distortion contribution to the phase space density.  Writing the spectral distortion as
\begin{equation}
\Delta f_\nu = \left( \frac{x_{2p}}{3x_{1s}} - e^{-h\nula/\Tr} \right)\Psi(y),
\end{equation}
Ref.~\cite{2008PhRvD..78b3001H} showed that in the absence of frequency diffusion the rescaled spectral distortion $\Psi(y)$ satisfied the equation
\begin{equation}
\frac{d\Psi}{dy} = \frac W{y^2} (e^y\Psi-1)
\end{equation}
in the time-steady approximation,
i.e. the same equation as occurs in the red wing.  [This is Eq.~(109) in Ref.~\cite{2008PhRvD..78b3001H}; the missing $y^2$ in that paper is
a typo.]  The inclusion of frequency diffusion proceeds exactly analogously to \S\ref{ss:red}, yielding
\begin{equation}
0 = \frac{d\Psi}{dy} - \frac W{y^2}(e^y\Psi-1) + S\frac d{dy}\left[ y^{-2}\left(\frac{d\Psi}{dy} + \Psi\right) \right].
\label{eq:psi-blue}
\end{equation}
The abundance of distortion photons in the blue wing is then
\begin{equation}
x_+ \approx \frac{8\pi\nula^2\Tr}{c^3\nH h}\left( \frac{x_{2p}}{3x_{1s}} - e^{-h\nula/\Tr} \right){\cal I}(W,S),
\label{eq:x-plus}
\end{equation}
where
\begin{equation}
{\cal I}(W,S) = \int_0^\infty \Psi(y)\,dy
\end{equation}
is a dimensionless integral.  Values of ${\cal I}(W,S)$ are computed according to the method in Appendix~\ref{app:sol} and plotted in 
Fig.~\ref{fig:chiI}.

\subsection{Implementation and results}

Following the approach of Ref.~\cite{2008PhRvD..78b3001H}, we first compute the Ly$\alpha$ transport parameters $W(z)$ and $S(z)$ for the pure MLA code 
with all two-photon transitions and scattering effects turned off.  We then turn on the analytic corrections in Ref.~\cite{2008PhRvD..78b3001H} 
associated with the stimulated $2s\rightarrow 1s$ decays and nonthermal absorption, two-photon decays from $n\ge 3$ levels, and Raman scattering.  The 
two-photon decays from $n\ge 3$ levels depended on the function $\chi(W)\equiv\Phi(y=-\infty|W)$ for the sub-Ly$\alpha$ decays (i.e. those in which 
both of the emitted photons have $\nu<\nula$, which usually means one photon emerges in the red damping wing of Ly$\alpha$) and ${\cal I}(W)$ for the 
super-Ly$\alpha$ decays (where one photon has $\nu>\nula$, usually in the blue damping wing of Ly$\alpha$).  We can account for scattering 
semianalytically by replacing $\chi(W)$ and ${\cal I}(W)$ with their generalized values $\chi(W,S)$ and ${\cal I}(W,S)$ derived here.  We may then 
compare the resulting recombination histories with and without Ly$\alpha$ scattering.

\begin{figure}
\includegraphics[angle=-90,width=3.3in]{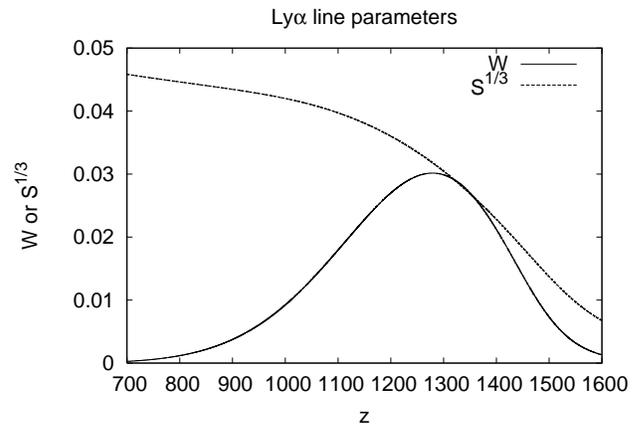}
\caption{\label{fig:ws}The dimensionless Ly$\alpha$ transport parameters $W(z)$ and $S(z)$.}
\end{figure}

The transport parameters $W(z)$ and $S(z)$ are shown in Fig.~\ref{fig:ws}.

The correction to the recombination history can be obtained by comparing the ``old'' $x_e(z)$ using $\chi(W)$ and ${\cal I}(W)$ without scattering to 
the ``new'' $x_e(z)$ using $\chi(W,S)$ and ${\cal I}(W,S)$.  The correction is shown by the dashed line in Fig.~\ref{fig:dcorr}.  Note the qualitative 
agreement with the fully numerical result, although at low redshifts our analytic approximation overestimates the correction.

\subsection{Comparison with other computations}

The changes in the recombination history that we have found amount to corrections of at most $0.45$\%.  This is less than the correction computed by 
several other authors.  This section discusses some possible explanations for the apparent discrepancies.  In some cases, the explanation lies with the 
fact that the corrections to the recombination history (or the effective escape probability) from emission/absorption in the damping wings are not 
additive with those from scattering -- the net effect of including both is less than one would expect from adding the contributions to $\Delta 
x_e/x_e$.  Also previous results on Ly$\alpha$ transfer did not include the deviation of emission versus absorption profiles, which can have a major 
impact on the results -- e.g. without this we would have $\chi(W,S=0)=1$ for any $W$.

Chluba \& Sunyaev \cite{2008arXiv0810.1045C} consider the time dependence of the radiation intensity in the Ly$\alpha$ line (i.e. 
non-quasi-stationarity) and the consequent effect on recombination.  They include true emission and absorption in the damping wings but not scattering, 
and so in terms of the physics their result is most comparable to the treatment of two-photon decays by Hirata \cite{2008PhRvD..78b3001H}.  In 
particular, Hirata found that the dominant time-dependent correction was that associated with the blue damping wing of Ly$\alpha$, i.e. with the time 
dependence of $\dot x_+$ (discussed here in \S\ref{ss:blue}).  Both Chluba \& Sunyaev \cite{2008arXiv0810.1045C} (see their Fig.~12) and Hirata 
\cite{2008PhRvD..78b3001H} (see his Fig.~8) find that the non-quasi-stationarity leads first to an accelerated recombination and then a delayed 
recombination as the spectral distortion redshifts through Ly$\alpha$, but Chluba \& Sunyaev find an effect up to a factor of $\sim 3$ larger.  We 
suspect this is due to their neglect of the deviation of emission versus absorption profiles, i.e. the $e^y$ factor in Eq.~(\ref{eq:psi-blue}).  
Without this factor (and with $S=0$) we derive the solution $\Psi(y)=1-e^{-W/y}$ and hence the integral ${\cal 
I}(W,S=0)=\int_0^\infty\Psi(y)dy=\infty$, and so the analytic approximation in Ref.~\cite{2008PhRvD..78b3001H} would yield an infinite correction in 
this approximation.  Chluba \& Sunyaev find a finite correction because of the finite duration of recombination (they use an exact treatment of 
non-quasi-stationarity rather than Hirata who treats the time dependence as a perturbation), but the effect is still large.

Grachev \& Dubrovich \cite{2008AstL...34..439G} computed a correction to Ly$\alpha$ escape based on the modified escape probability of Grachev 
\cite{1989Ap.....30..211G}.  The latter did not include non-time-steady effects and was based on the rate of redshifting of Ly$\alpha$ photons out of 
the line, similar to our \S\ref{ss:red}.  They also did not include the deviation of emission versus absorption profiles, which is equivalent to 
ignoring the $e^y$ in our Eq.~(\ref{eq:phi-red}).  This led them to the analytic approximation of Grachev \cite{1989Ap.....30..211G}, which is
\begin{equation}
\chi-1 = \rho_{\rm G} \left[ 1 + \frac{\sigma_{\rm G}^2(4-\sigma_{\rm G}^2)}{3(2+\sigma_{\rm G}^2)} + \frac{\sigma_{\rm G}^4}6
\right]^{-1},
\label{eq:grachev-chi}
\end{equation}
where Grachev's dimensionless parameters are $\sigma_{\rm G}^2 \equiv 3^{2/3}W/\sqrt[3]S$ and $\rho_{\rm G} \equiv \sqrt[3]{3S}$.
[Grachev \cite{1989Ap.....30..211G} uses $i(-\infty)$ in place of our $\chi$.]
We have integrated our equation by the method of Appendix~\ref{app:sol} without the $e^y$ factor and find that this agrees with
Eq.~(\ref{eq:grachev-chi}) to better than 10\% in the range of interest.
We note that our change in $\chi$ due to scattering, $\chi(W,S)-\chi(W,0)$, is increased if we turn 
off this term, in qualitative agreement with the fact that Grachev \& Dubrovich \cite{2008AstL...34..439G} find a larger change in the recombination 
history due to Ly$\alpha$ scattering.

\section{Implications for CMB anisotropies}
\label{sec:cmb}

\begin{figure}
\includegraphics[angle=-90,width=3.3in]{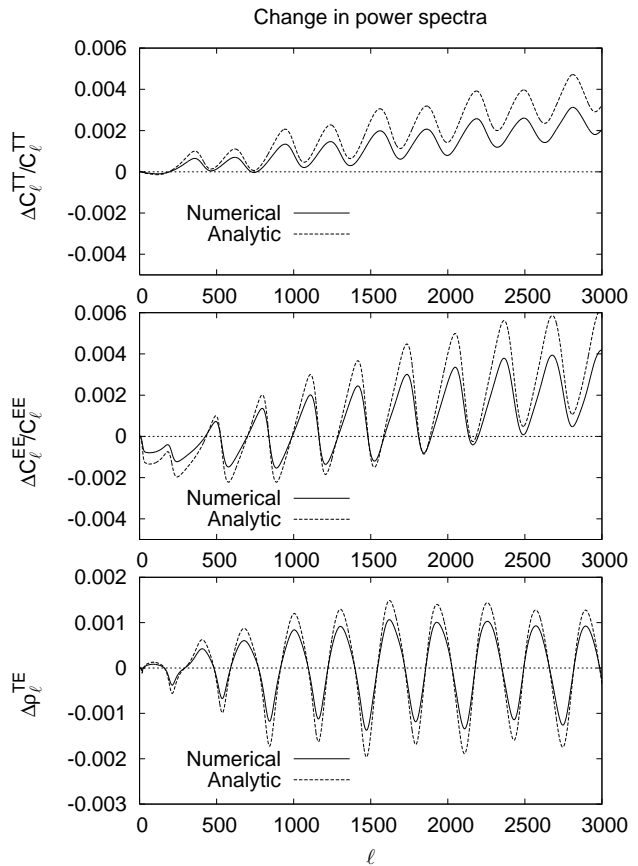}
\caption{\label{fig:sc}The change in CMB power spectra for the temperature (top panel), polarization (middle panel), and the correlation coefficient 
(bottom panel) due to inclusion of Ly$\alpha$ scattering.  The overall tilt and oscillating behavior are both the result of a faster recombination and 
hence higher redshift of the surface of last scattering: the tilt from the reduced Silk damping, and the oscillations due to the smaller acoustic 
horizon.}
\end{figure}

In Fig.~\ref{fig:sc}, we evaluate the effect of the Ly$\alpha$ diffusion correction on the CMB anisotropies, i.e.
\begin{equation}
\frac{\Delta C_\ell^{TT}}{C_\ell^{TT}} = \frac{C_\ell^{TT}({\rm with~scattering})}{C_\ell^{TT}({\rm without~scattering})}-1.
\end{equation}
Note that the two-photon transitions are turned on in both the ``with diffusion'' and ``no diffusion'' case.  The CMB power spectrum is computed using 
the Boltzmann code {\sc Cmbfast} \cite{1996ApJ...469..437S}.  The most obvious effect is the oscillation in $\Delta C_\ell/C_\ell$: these are due to a 
slight shift in the acoustic scale to higher $\ell$ since the faster recombination results in a higher-redshift surface of last scattering.  The 
overall tilt is partly a result of the reduced Silk damping: the faster recombination gives less time for the acoustic oscillations to be damped by 
photon diffusion, and hence the small-scale perturbations are not as suppressed as in the standard scenario.  Also the reduced electron density at 
$z\sim 900$ implies a lower optical depth after the surface of last scattering and hence less washing out of the small-scale features in the CMB.  This 
latter effect is slightly overestimated by the analytic computation, which is why the analytic result (dashed line in Fig.~\ref{fig:sc}) has a slightly 
larger high-$\ell$ power spectrum.

As in Ref.~\cite{2008PhRvD..78b3001H}, we may evaluate the importance of the Ly$\alpha$ scattering correction for a given experiment by considering
\begin{equation}
Z^2 = \sum_{\ell\ell'} F_{\ell\ell'} \Delta C_\ell^{TT} \Delta C_{\ell'}^{TT};
\end{equation}
here $F_{\ell\ell'}$ is the experiment's Fisher matrix and
$Z$ is the maximum number of sigmas by which any parameter fit could be affected.  (An individual cosmological parameter may or may not be affected, 
depending on whether its effect on the CMB power spectrum is similar to that caused by the inclusion of scattering.)  We consider the implications 
for the WMAP 2008 (5-year) data \cite{2009ApJS..180..330K} (including beam and point source errors), the Arcminute Cosmology Bolometer Array Receiver 
(ACBAR) 2008 power spectrum 
\cite{2008arXiv0801.1491R} (including beam and calibration errors), and the upcoming {\slshape Planck} data as forecast using the noise curves for the 
70 GHz (Low-Frequency Instrument) and 
100 and 143 GHz (High-Frequency Instrument) channels in the Blue Book \cite{2006astro.ph..4069T} and assuming a usable sky coverage of $f_{\rm 
sky}=0.7$.

Results of this analysis are displayed in Table~\ref{tab:correction}.  It is seen that the correction due to Ly$\alpha$ scattering is small -- it is 
only 0.9$\sigma$ for {\slshape Planck}.  Moreover, it goes in the opposite direction to the two-photon corrections found by 
Ref.~\cite{2008PhRvD..78b3001H} (the scattering raises the high-$\ell$ power spectrum whereas the two-photon corrections lower it).

\begin{table}
\caption{\label{tab:correction}The magnitude of the correction $Z$ (in number of sigmas) to the CMB power spectrum for several CMB experiments.  This 
is shown both for the Ly$\alpha$ diffusion correction (center column) and for the combined two-photon decay corrections \cite{2008PhRvD..78b3001H} and 
diffusion correction (right column).}
\begin{tabular}{lcccc}
\hline\hline
Experiment & & $Z$ & & $Z$ \\
 & & scattering & & 2$\gamma$+scattering \\
\hline
WMAP 5yr & & 0.07 & & 0.23 \\
ACBAR 2008 & & 0.04 & & 0.17 \\
{\slshape Planck} & & 0.92 & & 5.01 \\
\hline\hline
\end{tabular}
\end{table}

\section{Discussion}
\label{sec:conc}

We have examined the effect of multiple scattering on the \HI\ Ly$\alpha$ escape problem during the cosmological recombination epoch.  Both numerical 
and analytic tools were developed to treat this problem simultaneously with emission and absorption in the damping wings.  We find that scattering 
increases the escape probability and speeds up recombination because atomic recoil leads to a systematic shift of the photons to the red wing of the 
Ly$\alpha$ line.  While this is in qualitative agreement with previous results \cite{1990ApJ...353...21K, 2008AstL...34..439G}, the magnitude of the 
scattering correction is less than previously suggested.  We believe this is because the proper treatment of emission and absorption leads to a smaller 
effect from scattering.  The modified treatment of the \HI\ Ly$\alpha$ resonance results in a very small correction to the CMB power spectrum, ranging 
from zero to 0.4\% at $\ell<3000$, of the order of 0.9$\sigma$ for {\slshape Planck}.

Our ultimate goal in hydrogen recombination is to develop a complete theory with an error budget as has been done for helium 
\cite{2008PhRvD..77h3008S}.  This will be based in part on the work presented in this paper, but will also require investigation of many minor
radiative and collisional processes.  Eventually the theory must be encapsulated in a code fast enough to use in Markov chain parameter estimation
techniques, which could be the result of either analytic simplifications or interpolation codes \cite{2008arXiv0807.2577F}.

\begin{acknowledgments}

We thank Y. Ali-Ha\"imoud, D. Grin, and E. Switzer for useful discussions and comments.

This project was supported by the U.S. Department of Energy (DE-FG03-92-ER40701) and the National Science Foundation (AST-0807337).  C.H. is supported 
by the Alfred P. Sloan Foundation.  J.F. received support from Caltech's Summer Undergraduate Research Fellowship (SURF) program and the Flintridge 
Foundation.

\end{acknowledgments}

\appendix

\section{Solution to time-steady diffusion equation}
\label{app:sol}

In this appendix, we consider the solution to the dimensionless time-steady radiative transfer equation, Eq.~(\ref{eq:phi-red}).  In the red damping 
wing, $y<0$, we desire the value of the solution at large negative values $\Phi(-\infty)$.  In the blue wing, $y>0$, we desire the integral 
$\int_0^\infty\Phi(y)dy$.  Our ODE is 
different from the case of no frequency diffusion \cite{2008PhRvD..78b3001H} because the diffusion operator renders it second-order.  It can be solved 
by defining the variable
\begin{equation}
\Xi = Sy^{-2} \left( \frac{d\Phi}{dy} + \Phi \right).
\end{equation}
Here $\Xi(y)$ can be thought of as a dimensionless flux of photons passing through $y$ due to
frequency diffusion ($Sy^{-2}d\Phi/dy$) and recoil ($Sy^{-2}\Phi$).  It is exactly zero without the diffusion/recoil terms.
This leads to the linear system
\begin{eqnarray}
\frac{d\Phi}{dy} &=& \frac{y^2}S\Xi - \Phi,
\nonumber \\
\frac{d\Xi}{dy} &=& Wy^{-2}(e^y\Phi-1) - \frac{y^2}S\Xi + \Phi,
\label{eq:system}
\end{eqnarray}
which is singular at $y=0$.

We then implement a shooting method in the red damping wing, and a numerically stable modification of the shooting method in the blue wing.

\subsection{Red wing}

First consider the red wing, $y<0$.  The boundary conditions are that $\Phi(y=0)=1$ and $\Phi(y=-\infty)$ is finite.  In order for $\Phi(y=-\infty)$ to 
be finite, inspection of the first equation in Eq.~(\ref{eq:system}) shows that we must have $\Xi(y=-\infty)=0$.  Therefore given any 
choice of $\chi=\Phi(y=-\infty)$, we may construct a solution $\Phi(y|\chi)$ to Eq.~(\ref{eq:system}) by taking $\Phi=\chi$ and $\Xi=0$ at large 
negative $y$ and integrating toward $y=0$ using a second-order implicit ODE integrator.

The problem remains to choose the correct value of $\chi$.  In general for small negative $y$, the behavior of the solution is that $\Phi(y)$ remains 
finite because of the $y^2$ suppression in the first equation of Eq.~(\ref{eq:system}), while $\Xi(y)$ can have a $\sim y^{-1}$ divergence as one 
can see from the second equation:
\begin{equation}
\lim_{y\rightarrow 0^-}\Phi(y) = \alpha, \;\;\;\;
\Xi(y) = \frac{W(1-\alpha)}y + {\cal O}(y^0).
\end{equation}
The desired solution is that corresponding to the boundary condition $\alpha=1$.
As we integrate the solution $\Phi(y|\chi)$ toward $y=0$, we can determine the value $\alpha$ as a function of $\chi$.  Since the ODE is 
linear, $\alpha(\chi)$ is a linear function of $\chi$, and it suffices to obtain $\alpha(\chi=0)$ and $\alpha(\chi=1)$.  Then the desired value of 
$\chi$ that satisfies the line center boundary condition can be obtained by linear extrapolation:
\begin{equation}
\chi(\alpha=1) = \frac{1-\alpha(0)}{\alpha(1)-\alpha(0)}.
\end{equation}
Since the fundamental problem is to find $\Phi(y=-\infty)$, we may simply report $\chi(\alpha=1)$.

\subsection{Blue wing}

The solution in the blue wing, $y>0$, must satisfy the boundary conditions $\Phi(y=0)=1$ and $\Phi(y=+\infty)=0$.  As before, we will implement a 
shooting method in the $+y$ direction, i.e. starting at small positive $y=\epsilon$ and integrating toward $+\infty$.  But here the numerical 
calculation is much trickier: whereas in the red wing the growing mode of Eq.~(\ref{eq:system}) had only a power-law divergence ($\Xi\propto y^{-1}$) 
as one approached line center, here the growing mode grows faster than exponentially.  Therefore if we push to large positive $y$ an overflow error 
occurs.  More seriously, this growing mode implies that the correct initial condition $\Xi(y=\epsilon)$ cannot be well represented even in double 
precision.

A slow but effective way to solve this problem is to write a function that takes in an initial value of $y_{\rm i}$ and a dimensionless photon density 
$\Phi(y_{\rm i})$ at that point, and find the critical value $\Xi_{\rm c}[y_{\rm i},\Phi(y_{\rm i})]$ of $\Xi$ that satisfies the boundary condition at 
$y=+\infty$ (i.e. is 
non-divergent).  Here $\Xi_{\rm c}$ can be determined by a shooting method using a second-order implicit ODE integrator with two trials, taking 
advantage of linearity, just as we did for the red wing.  We also determine how far one can integrate before the dangerous growing mode kicks in.  This 
can be found from the function
\begin{equation}
F[y|y_{\rm i},\Phi(y_{\rm i})] = \left.\frac{\partial\Phi[y|y_{\rm i},\Phi(y_{\rm i}),\Xi(y_{\rm i})]}{\partial\Xi(y_{\rm i})}
\right|_{\Xi(y_{\rm i})=\Xi_{\rm c}}e^{y-y_{\rm i}}
,
\end{equation}
which takes the initial value $F(y_{\rm i})=0$ and rises toward $\infty$ as $y$ increases.  We solve for $y_{\rm th}[y_{\rm i},\Phi(y_{\rm i})]$ at 
which $F=10^4$, i.e. where the growing mode has incresed by 4 orders of magnitude.  The exponential $e^{y-y_{\rm i}}$ is included because the physical 
solution has $\sim e^{-y}$ dependence at large $y$.

We then begin our solution to Eq.~(\ref{eq:system}) by starting at $y=\epsilon$, $\Phi(y)=e^{-\epsilon}$, and 
$\Xi(y)=\Xi_{\rm c}(\epsilon,e^{-\epsilon})$.  
The implicit ODE integrator is used to integrate to larger $y$, but only until we reach $y_{\rm th}$.  Beyond this point the growing mode is dangerous, 
so we recompute the critical $\Xi_{\rm c}$ at $[y,\Phi(y)]$, take this as a new initial condition, and keep integrating until the new $y_{\rm th}$.  
This procedure ``resets'' the unstable mode each time a threshold value $y_{\rm th}$ is reached.

At very large values of $y$, where the growing mode grows very quickly, the above procedure becomes very slow.  Fortunately, we find that at this point 
$\Phi$ is well-represented by the large-$y$ asymptotic result $\Phi(y)\rightarrow e^{-y}$.  We may then extract the integral
${\cal I}(W,S) = \int_0^\infty \Phi(y|W,S)\,dy$.

In the real Universe, the solution at very large values of $y$ is not accurate because of the presence of other resonances (e.g. Ly$\beta$ is at $y\sim 
8$ at $z=1000$).  Fortunately, these large values of $y$ do not contribute significantly to the integral ${\cal I}(W,S)$ because of the $e^{-y}$ 
suppression of the integrand; it is for this reason that the analytic approximation based on ${\cal I}(W,S)$ works so well.

\bibliography{reclya}

\end{document}